\documentclass[]{article}

\usepackage{PRIMEarxiv}

\usepackage[utf8]{inputenc} 
\usepackage[T1]{fontenc}    
\usepackage{hyperref}       
\usepackage{url}            
\usepackage{booktabs}       
\usepackage{amsfonts}       
\usepackage{nicefrac}       
\usepackage{microtype}      
\usepackage{lipsum}
\usepackage{fancyhdr}       
\usepackage{graphicx}       
\graphicspath{{media/}}     

\usepackage{amssymb}
\usepackage{amsmath}
\usepackage{cleveref}
\usepackage{bbm}
\usepackage{algorithm}
\usepackage{algpseudocode}
\usepackage{apacite}
\usepackage[numbers]{natbib}
\title{Personalized Improvement of Standard Readout Error Mitigation using Low-Depth Circuits and Machine Learning
\thanks{\textit{\underline{Citation}}: 
\textbf{Lee. PEM}} 
}

\author{
  Melody Lee \\
  College of Computing  \\
  Georgia Institute of Technology \\
  Atlanta, Georgia\\
  \texttt{mlee769@gatech.edu}
}

\begin{document}
\maketitle

\begin{abstract}
Quantum computers have shown promise in improving algorithms in a variety of fields.
The realization of these advancements is limited by the presence of noise and high error rates, which become prominent especially with increasing system size.
Mitigation techniques using matrix inversions, unfolding, and deep learning, among others, have been leveraged to reduce this error.
However, these methods are not reflective of the entire gate set of the quantum device and may need further tuning depending on the distance from the most recent calibration time.
This paper proposes a method of improvement to numerical readout error techniques, where the readout error model is further refined using measured probability distributions from a collection of low-depth circuits.
We use machine learning to improve the readout error model for the quantum system, testing the circuits on the simulated IBM Perth backend.
We demonstrate a median 6.6\% improvement in fidelity, 29.9\% improvement for mean-squared error, and 10.3\% improvement in Hellinger distance over the standard error mitigation approach for a seven-qubit system with a circuit depth of four.
With further focus directed towards such improvement of these error mitigation techniques, we are one step closer to the fault-tolerant quantum computing era.
\end{abstract}


\section{Introduction}\label{s:introduction}
Quantum computers and information theory are projected to impact a broad range of scientific and industrial work~\cite{moller2017impact,de2017potential,mavroeidis2018impact,rawat2022quantum}. 
Noise, however, presents a significant challenge to realizing the full potential of this quantum technology~\cite{saki2021impact, resch2021benchmarking}.
Many near-term quantum devices become susceptible to error with scaling system size and depth.
This scalability of quantum devices is of particular interest in the current Noisy Intermediate-Scale Quantum (NISQ) era~\cite{preskill2019quantum}.
Presently, the ability of quantum advantage to be realized by these NISQ devices is under contention due in large part to such noise-based limitations~\cite{arute2019quantum, pan2022solving, farhi2014quantum, cerezo2021variational, stilck2021limitations, de2023limitations, aharonov2023polynomial}.
To address such concerns, efforts have been directed towards a number of algorithms and protocols that seek to leverage methods that fall within current technological capabilities.
This includes the advent of variational quantum algorithms, which use hybrid quantum-classical devices, and optimization-focused algorithms~\cite{lubasch2020variational, bittel2021training, bonet2023performance}.

This quantum-classical approach to computation has also been applied to error mitigation strategies.
In particular, this paper focuses on quantum readout error, which has been a subject of interest in prior bodies of work largely due to its persistence even after fault tolerance implementation~\cite{aasen2025mitigation, svastits2025readout, yang2022efficient}.
Early methods for readout error mitigation sought to implement post-measurement processing by modeling the error using statistical models.
The results from quantum circuits are obtained as probability distributions over some finite set of possible measurement outcomes.
Prior works represent the readout error using confusion matrices~\cite{maciejewski2020mitigation, nation2021scalable}, demonstrating methods to learn and invert the readout error noise model.
These techniques are effective and are applied in widely-used software, such as Qiskit~\cite{javadi2024quantum}.
However, these inversions are sensitive to statistical fluctuations.
Thus, Bayesian unfolding has been explored as an alternative to such an approach~\cite{nachman2020unfolding, pokharel2024scalable}.

However, an $n$-qubit system has $2^n$ possible states.
As such, representation and rigorous inversion of the matrix will require an exponential amount of time and memory, even with inversion heuristics.
The statistical mitigation approach can be simplified with the assumption of sparse probability distributions~\cite{yang2022efficient}, which assumes only a few dominant measurement outcomes.
Under such assumptions, methods such as parallelization~\cite{betancourt2007parallel} and the application of the scaled Lasso~\cite{sun2013sparse} with a relatively lower order are possible.
Others have sought to leverage deep learning and general machine learning techniques to identify the error mitigation matrices~\cite{lienhard2022deep,kim2022quantum, liao2024machine}.
These methods apply machine learning techniques, such as linear regression, multilayer perceptrons, and random forest models, to learn the entirety of the error mitigation model.

However, the probability distributions obtained using predefined quantum circuits tend only to be reflective of limited gate sets.
In \citet{liao2024machine}, for example, a circuit composed of CX and Hadamard gates is used to generate data for their machine learning for quantum error mitigation (ML-QEM) approach.
Quantum circuits are implemented with various gate sets, which are calibrated to predefined parameters~\cite{cerfontaine2020self, maksymov2021optimal, fan2025calibrating}.
To improve the learned readout errors to a personalized gate set and account for potential inconsistencies in calibration, tuning the readout error model to reflect the state of the quantum computer may improve mitigation results.

This paper seeks to introduce a method of improvement to quantum readout error mitigation via statistical methods.
We leverage machine learning to generalize the readout error matrix for a quantum gate set using low-depth, but diverse, quantum circuits.
The approach focuses on improvement of the predefined, learned readout error model.

This paper is arranged as follows.
\textit{Section \ref{s:methods}} outlines the approach to both standard error mitigation and the personalized modification.
Within the section, \textit{Section \ref{s:notation_and_probability}} defines the notation used in the paper, and \textit{Section \ref{s:quantum_readout_error}} briefly discusses the theoretical framework.
\textit{Section \ref{s:personalized_learning}} highlights the machine learning-based approach.
\textit{Section \ref{s:results}} describes the results from the outlined approach.
The conclusion and directions for further improvement and research are contained in \textit{Section \ref{s:discussion}}.

\section{Methods}\label{s:methods}
\subsection{Notation and probability}\label{s:notation_and_probability}
Quantum states can be represented using either bra-ket notation or a vector-matrix representation.
Assume that a quantum bit, or qubit, is measured such that the possible outcome for a single qubit is either zero or one.
Define the measurement basis, then, of a single qubit to be in the state space with basis vectors
\begin{equation}
    |0 \rangle = \begin{bmatrix}
    1 \\
    0 
    \end{bmatrix} \text{ and } |1 \rangle = \begin{bmatrix}
    0 \\
    1 
    \end{bmatrix} .
\end{equation}
The state of each qubit can be represented as a linear combination of this basis, such that a singular qubit in state $| \psi \rangle$ is denoted $|\psi \rangle = \alpha |0 \rangle + \beta |1 \rangle$.
Here, $\alpha, \beta \in \mathbb{C}$, with constraint that $|\alpha|^2 + |\beta|^2 = 1$.
Physically, the squared norms of the amplitudes $\alpha$ and $\beta$ are the probabilities that the measurement of the qubit in state will result in the corresponding value of zero or one, respectively.

Now, consider a multi-qubit quantum system of $n$ qubits.
Define shorthand notation $|j \rangle$, where $|j \rangle$ refers to the bit state whose base-10 value is $j$.
For example, for $n=3$ we have $|3 \rangle = |011 \rangle$.
We represent state $|j \rangle$ using one-hot encoded vectors, where the vector contains all zero entries except a one in the $j^{\text{th}}$ position.
By these definitions, the state space of the system is given by
\begin{equation}\label{e:state_space}
    \chi = \{|j \rangle \forall j \in 0, \dots, 2^n - 1\}
\end{equation}
where $\chi \subseteq \mathbb{C}^n$ is a Hilbert space with defined inner product. That is, for any state $| \psi \rangle \in \chi$, we have
\begin{align*}
    |\psi \rangle = \alpha_0 |0 \rangle + \alpha_1 |1 \rangle + \dots + \alpha_{2^n - 1} |2^n - 1 \rangle
        = \sum_{j = 0}^{2^n - 1} \alpha_j |j \rangle
        =  \begin{bmatrix}
    \alpha_0 \\
    \alpha_1 \\
    \dots \\
    \alpha_{2^n - 1} 
    \end{bmatrix} \label{e:coefficients}
\end{align*}
with normalized coefficients and constraint
\begin{equation}\label{e:normalized_coefficients}
    \langle \psi | \psi \rangle = \sum_{j = 0}^{2^n - 1} |\alpha_j|^2 = 1
\end{equation}
where notation dictates that $\langle \psi | = |\psi \rangle^T$.
Physically, the norm $|\alpha_j|^2$ determines the probability that a given state $|j \rangle$ is obtained when $|\psi \rangle$ is measured in the standard basis.

Let $P$ denote a real-valued function that indicates the probability of measuring a certain outcome.
We now define generalized quantum measurements.
Let $\{M_{j}\}$ be the set of positive operator-valued measures (POVM) such that the probability of measuring state $|\psi \rangle$ to be $|j \rangle$ can be expressed as
\begin{equation}
    P(|j \rangle) = \langle \psi | M_{j}^{\dagger}M_{j} |\psi \rangle.
\end{equation}

POVMs are a particular class of operators that yield the expected probability distribution $P$ over a series of measurements based on their expectation values.
As such, POVMs must adhere to the following properties:
\begin{enumerate}
    \item $M_{j}^{\dagger} = M_j$: That is, the complex conjugate must be equivalent to the initial operator. This guarantees the computed probability will be real.
    \item $M_{j} \geq 0$: This ensures the predicted probabilities are positive.
    \item $\sum_{j} M_j = I$: This guarantees that the total expectation values sum to unity.
\end{enumerate}
We assume that measurements are performed using these POVMs to capture both classical noise and coherent errors at readout.
We can relate these probabilities by measurement to the coefficients of a quantum state vector using the Born rule.
For any state $|j \rangle$, we have
\begin{equation}\label{e:p(j)=alpha^2}
    P(|j \rangle) = \langle j | \psi \rangle = |\alpha_j|^2.
\end{equation}
We denote the probability of measuring this same state $j$ in the noisy case using a horizontal line: $| \overline{j} \rangle$.

Thus, using \textbf{Equation \ref{e:p(j)=alpha^2}}, we define vector $p \in \mathbb{R}^{2^n}$, where the probability of obtaining state $|j \rangle$ is at index $j$, where
\begin{equation}\label{e:probability_vector}
    p = \begin{bmatrix}
        |\alpha_0|^2 \\
        |\alpha_1|^2 \\
        \vdots \\
        |\alpha_{2^n - 1}|^2 
        \end{bmatrix} = \begin{bmatrix}
        P(|0 \rangle) \\
        P(|1 \rangle) \\
        \vdots \\
        P(|2^n - 1 \rangle)
        \end{bmatrix}
\end{equation}
Let $\bar{p}$ refer to the probabilities obtained from repeated measurements $M$ on the results with error, such that, as with \textit{Equation \ref{e:probability_vector}}, we have
\begin{equation}\label{e:bar_probability_vector}
    \bar{p} = \begin{bmatrix}
        |\bar{\alpha}_0|^2 \\
        |\bar{\alpha}_1|^2 \\
        \vdots \\
        |\bar{\alpha}_{2^n - 1}|^2 
        \end{bmatrix} = \begin{bmatrix}
        P(|\bar{0} \rangle) \\
        P(|\bar{1} \rangle) \\
        \vdots \\
        P(|\bar{2^n - 1} \rangle)
        \end{bmatrix}
\end{equation}

\subsection{Quantum readout error}\label{s:quantum_readout_error}
Quantum readout error refers to the case when the state measured from the system is different from what was prepared.
For example, suppose the state $|1 \rangle$ is prepared on a single qubit.
A system with readout error could measure $|0\rangle$ from the prepared state with a nonzero probability.
When stated in terms of probability $P$, we have that $0 < P(|\overline{0} \rangle | |1 \rangle) \leq 1$, where $|\overline{0}\rangle$ is the measured outcome.

We employ the statistical model for readout error representation.
Let us define the readout error matrix for $n$ qubits to be
\begin{equation}\label{e:readout_error_matrix}
    \mathcal{E} = \begin{bmatrix}
        P(|\overline{0} \rangle | |0 \rangle) & P(|\overline{1} \rangle | |0 \rangle) & \cdots & P(|\overline{2^n - 1} \rangle | |0 \rangle) \\
        P(|\overline{0} \rangle| |1 \rangle) & P(|\overline{1} \rangle | |1 \rangle) &  & \\
        \vdots & & \ddots & \vdots\\
        P(|\overline{0} \rangle | |2^n - 1 \rangle) & \cdots &  & P(|\overline{2^n - 1} \rangle | |2^n - 1 \rangle) 
    \end{bmatrix},
\end{equation}
where the arrangement of elements resembles that of a confusion matrix.
Here, $P(|\overline{j} | |k \rangle)$ for a $n$-qubit system refers to the probability of measuring a state $|j \rangle$ given an intended state of $|k \rangle$ for all $j, k \in [0, \dots, 2^n - 1$.

We make several key assumptions.
\begin{enumerate}
    \item \textbf{Assumption 1}: We assume that $|\overline{j} \rangle$ and $|k \rangle$ are random variables.
    \item \textbf{Assumption 2}: We assume that the readout error probability of each qubit in the $n$-bit system is independent.
    \item \textbf{Assumption 3}: We assume the probability of $P(|j \rangle | |k\rangle)$ is nonzero, where $P(|j \rangle | |k\rangle) << P(|k \rangle | |k\rangle)$ for all $j \neq k$.
\end{enumerate}
Concerning \textbf{Assumption 2}, qubits in reality are susceptible to imperfect control operators, physical interactions with nearest neighbors, and entanglement-related errors.

For any given state $j$, we compute the marginal probability, with adjustments for readout error, using the summation
\begin{equation}\label{e:sum_probabilities}
     P(|\overline{j} \rangle) = \sum_{k = 0}^{2^n - 1} P(|\overline{j} \rangle | |k \rangle) P(|k \rangle).
\end{equation}
Note that $P(|k\rangle) = |\alpha_k|^2$ from \textbf{Equation \ref{e:p(j)=alpha^2}}.
Furthermore, using \textbf{Equation \ref{e:bar_probability_vector}} and \textbf{Equation \ref{e:sum_probabilities}}, we approximate $\bar{p}$ using
\begin{align*}
    P(|\bar{j}\rangle) \approx \sum_{k = 0}^{2^n - 1} P(|\overline{j} \rangle | |k \rangle) |\alpha_k|^2 \nonumber\\
    \Rightarrow |\bar{\alpha}_j|^2 \approx \sum_{k = 0}^{2^n - 1} P(|\overline{j} \rangle | |k \rangle) |\alpha_k|^2 \nonumber\\
    \Rightarrow \bar{p}_j \approx \sum_{k = 0}^{2^n - 1} P(|\overline{j} \rangle | |k \rangle) p_k \nonumber\\
    \Rightarrow \bar{p} \approx \mathcal{E}^T p 
\end{align*}\label{e:barp_ETp}

\subsection{Readout error matrix computation}\label{s:em_computation}
To approximate the complete readout error matrix on $n$ qubits, $\mathcal{E}$, we perform a series of operations using experimental results.
Let this set of results be defined by $\{x \}$, where $x$ is a binary bit string of $n$ bits representing the measurement result of the quantum system with state $|\psi \rangle$.
Let $x_{i, j}$ be the digit in the $i^{\text{th}}$ index of the $j^{\text{th}}$ sample taken.

Given \textbf{Assumption 2}, we first compute the readout error matrix $Q_i$ for each qubit $i$ separately, then use these matrices to construct $\mathcal{E}$ for the entire bit string sequence.

Define the error matrix for the $i^{\text{th}}$ individual qubit to be 
\begin{equation}\label{e:q}
    \mathcal{Q}_i = \begin{bmatrix}
            P(|\bar{0}\rangle||0\rangle) & P(|\bar{1}\rangle||0\rangle) \\
            P(|\bar{0}\rangle||1\rangle) & P(|\bar{1}\rangle||1\rangle) 
            \end{bmatrix}.
\end{equation}

\begin{figure}[!htb]
   \begin{minipage}{0.48\textwidth}
     \centering 
     \includegraphics[width=.7\linewidth]{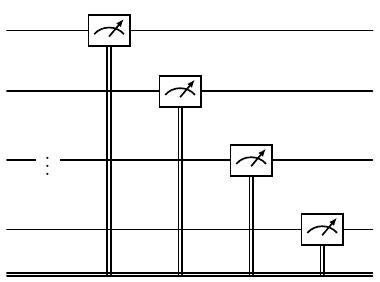}
     \caption{Circuit $\mathbb{O}$, which generates a quantum state that when measured has an expected value of $|00\dots0\rangle$. Used in initial readout error matrix construction.}\label{f:zeros}
   \end{minipage}\hfill
   \begin{minipage}{0.48\textwidth}
     \centering
     \includegraphics[width=.7\linewidth]{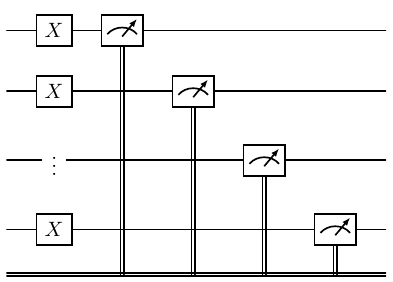}
     \caption{Circuit $\mathbbm{1}$, which generates a quantum state that, when measured, has an expected value of $|11\dots1\rangle$. Used in initial readout error matrix construction.}\label{f:ones}
   \end{minipage}
\end{figure}

To compute $P(|\bar{0}\rangle||0\rangle)$ and $P(|\bar{1}\rangle||0\rangle)$ for all $n$ qubits, we test a circuit $\mathbb{O}$, which has the expected result $|00\dots0\rangle$, on real hardware using $s$ shots. 
This circuit is illustrated in \textbf{Figure \ref{f:zeros}}.
For $\mathbb{O}$, we observe that for each qubit $i$, we have that
\begin{equation}\label{e:q_00}
        Q_{i, (0, 0)} = P(|\bar{0}\rangle |0\rangle) = \frac{|\{x_{i, j} = 0\}|}{s}
\end{equation}
and similarly
\begin{equation}\label{e:q_01}
    Q_{i, (0, 1)} = P(|\bar{1}\rangle |0\rangle) = \frac{|\{x_{i, j} = 1\}|}{s},
\end{equation}
where $Q_{i, (j, k)}$ refers to the entry in the $j^{\text{th}}$ row and $k^{\text{th}}$ column of matrix $Q_i$.

We compute $P(|\bar{0}\rangle||1\rangle)$ and $P(|\bar{1}\rangle||1\rangle)$ in a similar manner using test circuit $\mathbbm{1}$, as illustrated in \textbf{Figure \ref{f:ones}}.
This circuit has expected result $|11\dots1\rangle$ and, when run on real hardware using $s$ shots, we compute the probabilities for each qubit $i$
\begin{equation}\label{e:q_10}
    Q_{i, (1, 0)} = P(|\bar{0}\rangle |1\rangle) = \frac{|\{x_{i, j} = 0\}|}{s}
\end{equation}\label{e:q_11}
and
\begin{equation}
    Q_{i, (1, 1)} = P(|\bar{1}\rangle |1\rangle) = \frac{|\{x_{i, j} = 1\}|}{s}.
\end{equation}

Suppose we define binary strings $x, y \in \{x\}$ with decimal values of $j$ and $k$, where $x$ is the experimental binary string measured from the state $|\psi\rangle$ and $y$ is the expected binary string.
Then from the independence assumed in \textbf{Assumption 2}, we have
\begin{align*}
    \mathcal{E}_{j, k} = P(|\bar{j}\rangle | |k\rangle) \nonumber\\
        = \prod_{i=0}^{n-1} P(x_i = y_i) \nonumber\\
        = \prod_{i=0}^{n-1} Q_{i, (y_i, x_i)}\label{e:mitigation_matrix_entrywise}
\end{align*}
Thus, we compute the values of $\mathcal{E}$ entry-wise.
Since $\mathcal{E}$ is determined by obtaining the probability of a finite number of samples $s$, we expect a sampling error that scales inversely to the number of samples taken.

\subsection{Simple error mitigation}\label{s:em}
Post-measurement error mitigation aims to approximate $p$ using the measured $\bar{p}$.
We seek to compute some matrix $\mathcal{E}_m$ where $(\mathcal{E}\mathcal{E}_m)^T = \mathcal{E}_m^T\mathcal{E} \approx I$, such that when applied to both sides of \textbf{Equation \ref{e:barp_ETp}} we have
\begin{align*}
    \mathcal{E}_m^T\bar{p} \approx \mathcal{E}_m^T\mathcal{E}^Tp \nonumber\\
        \Rightarrow p \approx \mathcal{E}_m^T\bar{p}.
\end{align*}
From \textbf{Assumption 3}, we surmise
\begin{equation}
    \mathcal{E}_{k, k} >> \mathcal{E}_{j, k} > 0 \text{ for all } j \neq k.
\end{equation}
That is, $\mathcal{E}$ is diagonally dominant. 
Therefore, $\mathcal{E}$ is invertible and $\mathcal{E}^{-1}$ exists.
Let error mitigation matrix $\mathcal{E}_m = \mathcal{E}^{-1}$.
We refer to this method as the simple error mitigation (EM) approach. To compute $\mathcal{E}^{-1}$, we use a randomized lower-upper (LU) decomposition.

\subsection{Personalized learning}\label{s:personalized_learning}
To compute the personalized error mitigation (PEM) matrix, we first construct the approximate readout error model $\mathcal{E}$ using circuits $\mathbb{O}$ and $\mathbbm{1}$.
We intend to update this readout error model with further modifications to the final mitigation method.
Given $n$ qubits, to guarantee all $2^n$ possible measurement outcomes are sampled a sufficient number of times to reflect a trend in the backend behavior, an exponentially increasing number of quantum circuits in the training set is needed.
Instead, we use low-depth circuits to update the qubit-specific probability distribution matrices, $Q_i$, and construct an updated $\mathcal{E}^*$ using these results.

We first compute the initial $Q_i$ for the $i^{\text{th}}$ qubit, as outlined in \textit{Section \ref{s:em_computation}}.
Let a training set $\mathbb{T}$ of $N$ quantum circuits on $n$ qubits with depth $D$ be defined.
Define $q_{i, k} \in \mathbb{R}^2$ be a vector of measurement outcome probabilities for the $i^{\text{th}}$ qubit on the $k^{\text{th}}$ sample, such that we have
\begin{equation}
    q_{i, k} = \begin{bmatrix}
P(|0\rangle)\\
P(|1\rangle) 
\end{bmatrix},
\end{equation}
and $\bar{q}_{i, k}$ is the vector of probabilities for the same qubit and sample in the presence of error.
Then, given some $Q_i$ as defined in \textbf{Equation \ref{e:q}}, consider some qubit $i$ and sample $k$.
We find
\begin{align*}
    Q_i^T \cdot q_{i, k} = \begin{bmatrix}
        P(|\bar{0}\rangle | |0 \rangle) & P(|\bar{0}\rangle | |1 \rangle) \\
        P(|\bar{1}\rangle | |0 \rangle) & P(|\bar{1}\rangle | |1 \rangle) 
        \end{bmatrix}
        \begin{bmatrix}
        P(|0\rangle)\\
        P(|1\rangle) 
        \end{bmatrix} \nonumber\\
        = \begin{bmatrix}
        P(|\bar{0}\rangle | |0 \rangle) P(|0\rangle) + P(|\bar{0}\rangle | |1 \rangle) P(|1\rangle)\\
        P(|\bar{1}\rangle | |0 \rangle) P(|0\rangle) + P(|\bar{1}\rangle | |1 \rangle)  P(|1\rangle) 
        \end{bmatrix} \nonumber\\
        = \begin{bmatrix}
        P(|\bar{0}\rangle, |0\rangle) + P(|\bar{0}\rangle, |1\rangle) \\
        P(|\bar{1}\rangle, |0\rangle) + P(|\bar{1}\rangle, |1\rangle)
        \end{bmatrix} \nonumber\\
        = \begin{bmatrix}
        P(|\bar{0}\rangle)\\
        P(|\bar{1}\rangle) 
        \end{bmatrix},
\end{align*}
where the final term is computed by finding the marginal probabilities of $|\bar{0}\rangle$ and $|\bar{1}\rangle$.
Therefore,
\begin{equation}
    \bar{q}_{i, k} = Q_i^T \cdot q_{i, k}
\end{equation}
We seek to update the values in $Q_i$ using ordinary least squares linear regression.
For some qubit $i$, let $X_i \in \mathbb{R}^{N \times 2}$ be
\begin{equation}
    X_i = \begin{bmatrix}
q_{i, 0} \\
q_{i, 1} \\
\dots \\
q_{i, N} 
\end{bmatrix}.
\end{equation}
Suppose we seek to update the entries in $Q_i$ with measured result $|m\rangle$ for $m \in \{0, 1\}$.
Let
\begin{equation}
    \bar{y}_{i, m} = \begin{bmatrix}
\bar{q}_{i, 0, m} \\
\bar{q}_{i, 1, m} \\
\dots \\
\bar{q}_{i, N, m} 
\end{bmatrix}
\end{equation}
be the target, where we seek to optimize the least squares error by adjusting the weight vector $w_{i, m} \in \mathbb{R}^2$.
That is, given $X_i$, we seek to find weights $w_{i, m}$
\begin{equation}
    \bar{y}_{i, m} \approx X_i \cdot w_{i, m}
\end{equation}
such that the least squares error
\begin{equation}
    \epsilon = (\bar{y}_{i, m} - \hat{\bar{y}}_{i, m})^2,
\end{equation}
is minimized, where $\hat{\bar{y}}_{i, m} = X_i w_{i, m}$.
If $Q_i$ is a perfect representation of the error across all experiments, then $w_{i, m} = [Q_{i, (m, 0)}, Q_{i, (m, 1)}]^T$ and $\epsilon = \Vec{0}$.

Since the number of samples $N$ is greater than the two parameters being estimated for each $m \in \{0, 1\}$, and $X_i$ is nonzero, there exists closed form solution
\begin{equation}
    w_{i, m} = (X_i^T X_i)^{-1} X_i^T \bar{y}_{i, m}
\end{equation}
to the problem.
To update the initial $Q_i$, we define parameter $\eta \in (0, 1)$ as the learning rate, such that
\begin{equation}\label{e:eta}
    Q_i^* = c * ((1 - \eta) Q_i + \eta [w_{i, 0}, w_{i, 1}]),
\end{equation}
where $c$ is some normalization constant.

Given the training set, we compute the updated $Q_i^*$ for all qubits $i$.
Using these updated values, we compute an updated readout error matrix $\mathcal{E}^*$.
Note that since $\eta$ is nonzero, the updated $Q_i$ will be nonzero.
Since the weights $w_{i, m}$ are obtained from experimental results, we assume that \textbf{Assumption 3} holds.
From the computations for $\mathcal{E}^*$, we observe that this results in a diagonally dominant and invertible matrix.
Thus, we invert $\mathcal{E}^*$ by first finding the LU decomposition and inverting the subsequent matrices to compute the personalized error mitigation matrix $(\mathcal{E}^*)^{-1}$.

Then, to mitigate an experimental probability $\bar{p}$, we compute $((\mathcal{E}^{*})^{-1})^T \bar{p}$.
To account for possible negative entries in $w_{i, m}$ and, in turn, in $(\mathcal{E}^{*})^{-1}$, we drop small-magnitude, negative entries in the final probability vector and normalize the remaining values to find the final, mitigated probability distribution.

\subsection{Hardware and packages}\label{s:hardware_and_packages}
We use Qiskit, a quantum software stack built by IBM, for circuit generation and testing.
Expected probabilities are simulated on classical computers and obtained via the Qiskit Aer Simulator.
The test and training datasets are randomly generated with a predefined depth and number of qubits.
The experimental results with noise are obtained via circuits run on the Fake Perth backend, which mimics the noisy behavior of the IBM Perth superconducting quantum computer.
To perform randomized LU decomposition, we use the SciPy package.
The linear regression is performed using models from the Scikit-learn package.
Remaining computations and data processing are conducted using NumPy.
All code is written using Python.

\section{Results}\label{s:results}
\subsection{Error metrics}\label{s:error_metrics}
To evaluate the effectiveness of the method, we use three error metrics: (1) state fidelity, (2) mean-squared error, and (3) the Hellinger distance.
For two probability vectors $p$ and $q$, we define
\begin{equation}\label{e:fidelity_eq}
    \text{fidelity} = (\sum_{j = 0}^{2^n - 1} \sqrt{p_j q_j})^2.
\end{equation}
Fidelity indicates the probability that a given quantum state, when measured, may be identified as another, where a fidelity of zero indicates no correlation, while a fidelity of one indicates that $p$ is equivalent to $q$.
We use the mitigated probabilities obtained using both the EM and PEM methods to compute the fidelity against the expected probability distribution $p$.
The mean-squared error (MSE) for expected and noisy probability vectors $p$ and $\bar{p}$, respectively, is given by
\begin{equation}\label{e:mse_eq}
    \text{MSE} = \frac{1}{2^n - 1}\sum_{j=0}^{2^n - 1} (p_i - \bar{p}_i)^2.
\end{equation}
This metric aligns with the objective function defined in the linear regression for personalization of the error matrix.
As the name implies, the metric quantifies the absolute distance between the two vectors.
Therefore, a low MSE when comparing some mitigated $\bar{p}$ to the expected $p$ is ideal.
Lastly, we use the Hellinger distance to express the similarity between two probability distributions.
Given probability vectors $p$ and $q$, we use the discrete formula evaluation, where
\begin{equation}\label{e:hellinger_eq}
    \text{Hellinger distance} = \frac{1}{\sqrt{2}} \sqrt{\sum_{j=0}^{2n - 1}(\sqrt{p_i} - \sqrt{q_i})^2}
\end{equation}

We apply these formulae to analyze the performance of standard EM and PEM.

\subsection{Parameter Tuning}\label{s:parameter_tuning}
\begin{figure}[!htb]
    \begin{center}
        \includegraphics{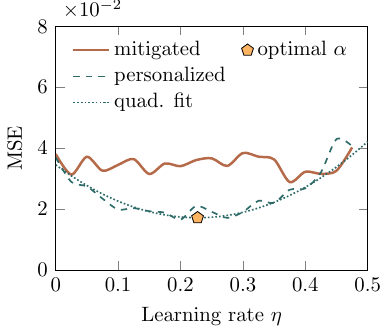}
        \caption{Identifying the learning rate $\eta$ by testing values from $0$ to $0.5$. The results have a quadratic best fit indicated in \textbf{Equation \ref{e:quadratic_fit}} and an optimal $\eta = 0.23$.}\label{f:tuning}
    \end{center}
\end{figure}

We first identify an optimal $\eta$, as defined in \textbf{Equation \ref{e:eta}} using hyperparameter tuning techniques.
We test $\eta$ values between zero and $0.5$, with a pre-defined data set consisting of $400$ training circuits and $50$ test circuits.
We evaluate the resulting error and perform a quadratic fit on the resulting MSE values to approximate the value for $\eta$ that minimizes the function, given by
\begin{equation}\label{e:quadratic_fit}
    \text{MSE} \approx 0.33\eta^2 - 0.15\eta + 0.035.
\end{equation}
The results are depicted in \textbf{Figure \ref{f:tuning}}, with the optimal $\eta = 0.23$ and a corresponding minimum MSE value of $0.02$.
We use this value of $\eta$ to obtain the remaining results, as discussed in \textit{Section \ref{s:mitigated_results}}.

\subsection{Mitigated Results}\label{s:mitigated_results}
Using the selected $\eta$ value, we construct standard EM and PEM matrices for seven qubits for the IBM Perth backend.
The PEM matrix is trained and personalized using a dataset of $1,000$ quantum circuits with a depth of four.

\begin{figure}[!htb]
    \begin{center}
        \includegraphics[width=.95\linewidth]{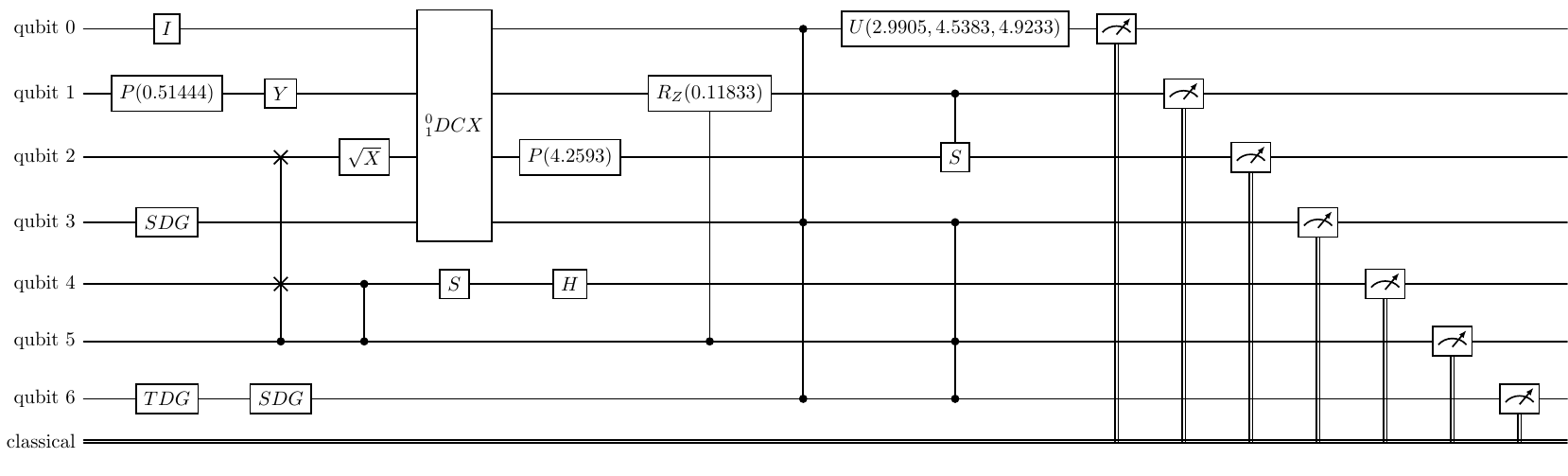}
        \caption{Example of a randomly generated circuit on $n=7$ qubits with global depth $D = 4$.}\label{f:example_circuit}
    \end{center}
\end{figure}

\begin{figure}[!htb]
    \begin{center}
        \includegraphics[width=0.95\linewidth]{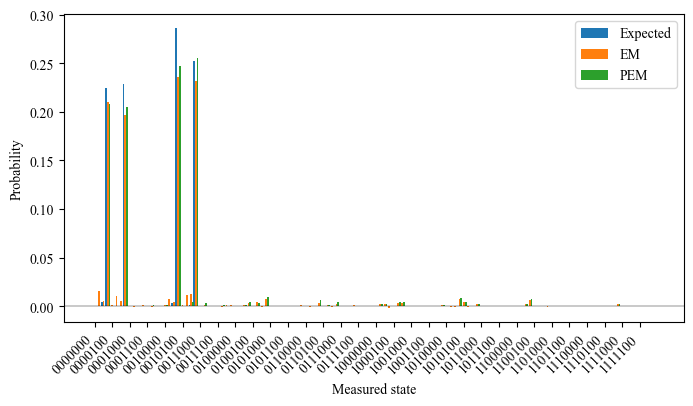}
        \caption{Comparison of the mitigated probabilities of measuring various quantum states, arranged in increasing order and compared against the expected probability distribution. The axis labels contain ticks for multiples of $2^3$ for the sake of readability.}\label{f:example_circuit_outcome}
    \end{center}
\end{figure}

We first test the obtained PEM matrix using an example circuit, illustrated in \textbf{Figure \ref{f:example_circuit}}, to demonstrate the viability of the method.
This circuit was randomly selected, again with parameters $n = 7$ and depth $D = 4$.
The probability distributions of the expected circuit behavior, EM method, and PEM method are illustrated in \textbf{Figure \ref{f:example_circuit_outcome}}.
Using the formulas defined in \textbf{Equation \ref{e:fidelity_eq}}, \textbf{Equation \ref{e:mse_eq}}, and \textbf{Equation \ref{e:hellinger_eq}}, we compute the fidelities for the different mitigation methods, as listed in \textbf{Table \ref{t:example_circuit}}.
Notably, we achieve a 49.3\% improvement in MSE over standard EM, with improvements across all three metrics.

\begin{table}[htp!]
\begin{center}
    \caption{Mitigation results for the circuit shown in \textbf{Figure \ref{f:example_circuit}}.}
    \begin{tabular}{|c|c|c|c|}
    \hline 
    \textbf{Method}         & \textbf{Fidelity}  & \textbf{MSE}          & \textbf{Hellinger Dist.} \\ \hline
    \textbf{Error mitigation (EM)}    & 0.887 & $0.005.32 \times 10^{-3}$ & 0.263 \\
    \hline
    \textbf{Personalized EM (PEM)}   & 0.914 & $0.002.70 \times 10^{-3}$ & 0.209\\ 
    \hline
    \textbf{\% Improvement} & 3.1\%              & 49.3\%                & 20.4\%  \\  
    \hline
    \end{tabular}
    \label{t:example_circuit}
\end{center}
\end{table}

\begin{figure}[!htp]
   \begin{minipage}{0.32\textwidth}
     \centering
     \includegraphics[width=.9\linewidth]{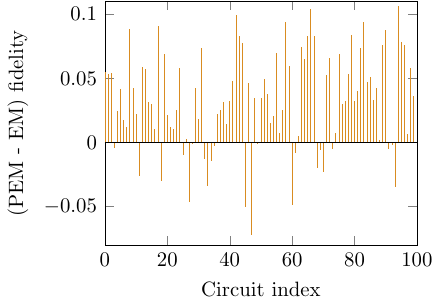}
   \end{minipage}\hfill
   \begin{minipage}{0.32\textwidth}
     \centering
     \includegraphics[width=.9\linewidth]{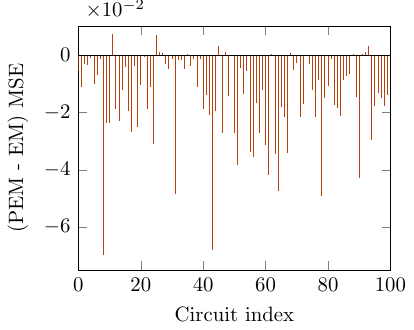}
   \end{minipage}
   \begin{minipage}{0.32\textwidth}
     \centering
     \includegraphics[width=.9\linewidth]{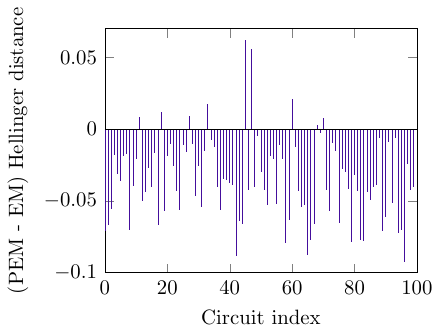}
   \end{minipage}
   \caption{The difference in evaluated error metric between the standard error mitigation and proposed PEM methods. From left to right, the metrics used are state fidelity, MSE, and Hellinger distance.}\label{f:metrics_difference}
\end{figure}

We then test the computed EM and PEM matrices on a randomized set of $100$ quantum circuits, which, like the example circuit, are randomly selected with parameters $n=7$ qubits and depth $D = 4$.
Over this larger dataset, we achieve a median 6.6\% improvement in fidelity, a 29.9\% improvement in MSE, and a 10.3\% improvement in the Hellinger distance by personalizing the EM matrix, as illustrated in \textbf{Table \ref{t:test_circuits}}.
The actual improvements for each test circuit are shown in \textbf{Figure \ref{f:metrics_difference}} for each metric, calculated using the difference between the quantitative metrics obtained for the PEM and EM methods.

Of the 100 circuits tested, 78\% demonstrate an improvement in fidelity, 85\% have an improvement in MSE, and 91\% experience an improvement in the calculated Hellinger distance when the PEM method is used over standard EM.
While the read-out probability distributions of most circuits are improved using PEM with promising probabilities, there exists a nontrivial number of circuits for which EM proves to be more effective for the metrics used here.
Given the randomized nature of the training data and test circuits, this is not wholly unexpected.
However, this indicates room for further improvement.

\begin{table}[htp!]
\caption{The proposed PEM method outperforms the simple EM method on all three indicated metrics for a test data set of $100$ randomly generated quantum circuits.}\label{t:test_circuits}
\begin{center}
    \begin{tabular}{|c|c|c|c|}
    \hline
    \textbf{Method}         & \textbf{Median Fidelity} & \textbf{Median MSE} & \textbf{Median Hellinger Dist.} \\ \hline
    \textbf{Error mitigation (EM)}   & 0.697                    & 0.047               & 0.414                           \\
     \hline
    \textbf{Personalized EM (PEM)}   & 0.743                    & 0.033               & 0.372                           \\ \hline
    \textbf{\% Improvement} & 6.6\%                    & 29.9\%              & 10.3\%     \\
     \hline
    \end{tabular}
\end{center}
\end{table}

\begin{figure}[!htb]
    \begin{center}
        \includegraphics[]{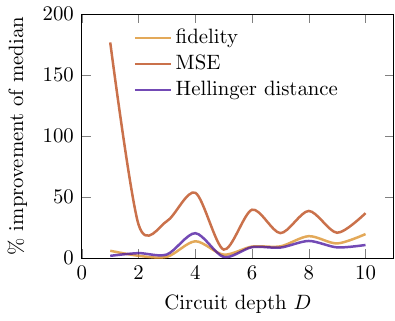}
        \caption{Performance of the PEM method as the depth of the circuit scales.}\label{f:depths}
    \end{center}
\end{figure}

We then demonstrate the performance of the PEM method over a range of circuit depths in \textbf{Figure \ref{f:depths}}.
The experiments train error mitigation matrices for varying depths from one to ten layers, using $400$ training samples and $50$ test samples.
The mitigation matrices assume $n=7$ total qubits are used on the IBM Perth backend.
\textbf{Figure \ref{f:depths}} shows that the improvement over the EM method appears to fluctuate as the circuit depth increases, but the PEM method maintains an advantage over its standard EM counterpart.

\section{Discussion}\label{s:discussion}
Within this paper, we present an improvement to the standard readout error mitigation method.
This improved method, PEM, personalizes the readout error model based on experimental results obtained from a training set of quantum circuits.
This personalization tunes qubit-specific probability distributions using machine learning.
These probability distributions are then used to construct an updated readout error model and, in turn, a personalized error mitigation protocol.
We evaluated the method using state fidelity, MSE, and the Hellinger distance as metrics, demonstrating an improvement over the standard EM approach for all three.
The employment of diverse, low-depth circuits for training allows for calibration of the error mitigation at regular intervals.

However, the personalization relies on a large training data set of quantum circuits.
For each of these circuits, the obtainment of a test probability distribution in the presence of noise requires repeated sampling.
While this is effective in sparse probability distributions, dense distributions across increasing qubit counts will require exponentially more samples to obtain probability estimates for all possible states.
By inspection, many of the randomly generated circuits in the test datasets have sparse probability distributions.
The post-processing error mitigation protocol, in which negative predicted values are dropped, is contingent on this assumption.
As such, further study of the effectiveness of the personalization method for more dense distributions is of interest.
Improvements for better performance for these cases are still needed.
To investigate this, future work may include the construction of custom sets of quantum circuits to maximize the coverage of possible gates and probability distributions.

The approach focuses on improvements specific to readout error mitigation.
Due to the nature of the measurement method used here, the probabilities reflect only the magnitudes of the coefficients for each quantum state.
Thus, the approach is effective only for quantum circuits whose usefulness is derived from these magnitudes, including circuits that prepare states with positive, real-valued coefficients.
While more complete state reconstruction methods to reflect the signs or imaginary values of state coefficients are possible, they scale exponentially with system size.
When coupled with the need for repeated sampling, this results in rapid growth in complexity, making it unfeasible for larger systems.
Nonetheless, further study of how the regression-based approach used in personalization can be applied to tune other mitigation strategies remains possible.

Additionally, the readout error model construction process itself scales rapidly with the number of qubits.
The computation of this matrix requires iteration through each entry.
Calculation for these entries requires repeated multiplications across all of the qubits in the system.
While this can be expressed as a series of matrix multiplications, which the NumPy package enables parallelization, the number of entries increases exponentially since the number of possible measurement outcomes per prepared state doubles with every additional qubit.
For vast numbers of qubits, the readout error matrix quickly grows to dimensions beyond current classical storage capabilities, presenting a computational limitation.
As the number of logical qubits available increases with the development of more robust quantum technologies, the future consideration of more efficient post-processing and computation approaches is necessary.

Nonetheless, this paper demonstrates that the use of experimental results can improve pre-existing numerical error mitigation strategies.
As the field of quantum information science and computing continues to develop, further exploration of these improvements is of great interest, setting the stage for fully realizing the capabilities of quantum computers.

\bibliography{references}

\begin{thebibliography}{}

\bibitem [\protect \citeauthoryear {%
Aasen%
, Di~Giovanni%
, Rotzinger%
, Ustinov%
\BCBL {}\ \BBA {} G\"{a}rttner%
}{%
Aasen%
\ \protect \BOthers {.}}{%
{\protect \APACyear {2025}}%
}]{%
aasen2025mitigation}
\APACinsertmetastar {%
aasen2025mitigation}%
\begin{APACrefauthors}%
Aasen, A\BPBI S.%
, Di~Giovanni, A.%
, Rotzinger, H.%
, Ustinov, A\BPBI V.%
\BCBL {}\ \BBA {} G\"{a}rttner, M.%
\end{APACrefauthors}%
\unskip\
\newblock
\APACrefYearMonthDay{2025}{}{}.
\newblock
{\BBOQ}\APACrefatitle {Mitigation of correlated readout errors without randomized measurements} {Mitigation of correlated readout errors without randomized measurements}.{\BBCQ}
\newblock
\APACjournalVolNumPages{arXiv preprint arXiv:2503.24276}{}{}{}.
\PrintBackRefs{\CurrentBib}

\bibitem [\protect \citeauthoryear {%
Aharonov%
, Gao%
, Landau%
, Liu%
\BCBL {}\ \BBA {} Vazirani%
}{%
Aharonov%
\ \protect \BOthers {.}}{%
{\protect \APACyear {2023}}%
}]{%
aharonov2023polynomial}
\APACinsertmetastar {%
aharonov2023polynomial}%
\begin{APACrefauthors}%
Aharonov, D.%
, Gao, X.%
, Landau, Z.%
, Liu, Y.%
\BCBL {}\ \BBA {} Vazirani, U.%
\end{APACrefauthors}%
\unskip\
\newblock
\APACrefYearMonthDay{2023}{}{}.
\newblock
{\BBOQ}\APACrefatitle {A polynomial-time classical algorithm for noisy random circuit sampling} {A polynomial-time classical algorithm for noisy random circuit sampling}.{\BBCQ}
\newblock
\BIn{} \APACrefbtitle {Proceedings of the 55th Annual ACM Symposium on Theory of Computing} {Proceedings of the 55th annual acm symposium on theory of computing}\ (\BPGS\ 945--957).
\PrintBackRefs{\CurrentBib}

\bibitem [\protect \citeauthoryear {%
Arute%
\ \protect \BOthers {.}}{%
Arute%
\ \protect \BOthers {.}}{%
{\protect \APACyear {2019}}%
}]{%
arute2019quantum}
\APACinsertmetastar {%
arute2019quantum}%
\begin{APACrefauthors}%
Arute, F.%
, Arya, K.%
, Babbush, R.%
, Bacon, D.%
, Bardin, J\BPBI C.%
, Barends, R.%
\BDBL {}others%
\end{APACrefauthors}%
\unskip\
\newblock
\APACrefYearMonthDay{2019}{}{}.
\newblock
{\BBOQ}\APACrefatitle {Quantum supremacy using a programmable superconducting processor} {Quantum supremacy using a programmable superconducting processor}.{\BBCQ}
\newblock
\APACjournalVolNumPages{Nature}{574}{7779}{505--510}.
\PrintBackRefs{\CurrentBib}

\bibitem [\protect \citeauthoryear {%
Betancourt%
\ \BBA {} Alvarado%
}{%
Betancourt%
\ \BBA {} Alvarado%
}{%
{\protect \APACyear {2007}}%
}]{%
betancourt2007parallel}
\APACinsertmetastar {%
betancourt2007parallel}%
\begin{APACrefauthors}%
Betancourt, R.%
\BCBT {}\ \BBA {} Alvarado, F\BPBI L.%
\end{APACrefauthors}%
\unskip\
\newblock
\APACrefYearMonthDay{2007}{}{}.
\newblock
{\BBOQ}\APACrefatitle {Parallel inversion of sparse matrices.} {Parallel inversion of sparse matrices.}{\BBCQ}
\newblock
\APACjournalVolNumPages{IEEE transactions on power systems}{1}{1}{74--81}.
\PrintBackRefs{\CurrentBib}

\bibitem [\protect \citeauthoryear {%
Bittel%
\ \BBA {} Kliesch%
}{%
Bittel%
\ \BBA {} Kliesch%
}{%
{\protect \APACyear {2021}}%
}]{%
bittel2021training}
\APACinsertmetastar {%
bittel2021training}%
\begin{APACrefauthors}%
Bittel, L.%
\BCBT {}\ \BBA {} Kliesch, M.%
\end{APACrefauthors}%
\unskip\
\newblock
\APACrefYearMonthDay{2021}{}{}.
\newblock
{\BBOQ}\APACrefatitle {Training variational quantum algorithms is NP-hard} {Training variational quantum algorithms is np-hard}.{\BBCQ}
\newblock
\APACjournalVolNumPages{Physical review letters}{127}{12}{120502}.
\PrintBackRefs{\CurrentBib}

\bibitem [\protect \citeauthoryear {%
Bonet-Monroig%
\ \protect \BOthers {.}}{%
Bonet-Monroig%
\ \protect \BOthers {.}}{%
{\protect \APACyear {2023}}%
}]{%
bonet2023performance}
\APACinsertmetastar {%
bonet2023performance}%
\begin{APACrefauthors}%
Bonet-Monroig, X.%
, Wang, H.%
, Vermetten, D.%
, Senjean, B.%
, Moussa, C.%
, B{\"a}ck, T.%
\BDBL {}O'Brien, T\BPBI E.%
\end{APACrefauthors}%
\unskip\
\newblock
\APACrefYearMonthDay{2023}{}{}.
\newblock
{\BBOQ}\APACrefatitle {Performance comparison of optimization methods on variational quantum algorithms} {Performance comparison of optimization methods on variational quantum algorithms}.{\BBCQ}
\newblock
\APACjournalVolNumPages{Physical Review A}{107}{3}{032407}.
\PrintBackRefs{\CurrentBib}

\bibitem [\protect \citeauthoryear {%
Cerezo%
\ \protect \BOthers {.}}{%
Cerezo%
\ \protect \BOthers {.}}{%
{\protect \APACyear {2021}}%
}]{%
cerezo2021variational}
\APACinsertmetastar {%
cerezo2021variational}%
\begin{APACrefauthors}%
Cerezo, M.%
, Arrasmith, A.%
, Babbush, R.%
, Benjamin, S\BPBI C.%
, Endo, S.%
, Fujii, K.%
\BDBL {}others%
\end{APACrefauthors}%
\unskip\
\newblock
\APACrefYearMonthDay{2021}{}{}.
\newblock
{\BBOQ}\APACrefatitle {Variational quantum algorithms} {Variational quantum algorithms}.{\BBCQ}
\newblock
\APACjournalVolNumPages{Nature Reviews Physics}{3}{9}{625--644}.
\PrintBackRefs{\CurrentBib}

\bibitem [\protect \citeauthoryear {%
Cerfontaine%
, Otten%
\BCBL {}\ \BBA {} Bluhm%
}{%
Cerfontaine%
\ \protect \BOthers {.}}{%
{\protect \APACyear {2020}}%
}]{%
cerfontaine2020self}
\APACinsertmetastar {%
cerfontaine2020self}%
\begin{APACrefauthors}%
Cerfontaine, P.%
, Otten, R.%
\BCBL {}\ \BBA {} Bluhm, H.%
\end{APACrefauthors}%
\unskip\
\newblock
\APACrefYearMonthDay{2020}{}{}.
\newblock
{\BBOQ}\APACrefatitle {Self-consistent calibration of quantum-gate sets} {Self-consistent calibration of quantum-gate sets}.{\BBCQ}
\newblock
\APACjournalVolNumPages{Physical Review Applied}{13}{4}{044071}.
\PrintBackRefs{\CurrentBib}

\bibitem [\protect \citeauthoryear {%
De~Palma%
, Marvian%
, Rouz{\'e}%
\BCBL {}\ \BBA {} Fran{\c{c}}a%
}{%
De~Palma%
\ \protect \BOthers {.}}{%
{\protect \APACyear {2023}}%
}]{%
de2023limitations}
\APACinsertmetastar {%
de2023limitations}%
\begin{APACrefauthors}%
De~Palma, G.%
, Marvian, M.%
, Rouz{\'e}, C.%
\BCBL {}\ \BBA {} Fran{\c{c}}a, D\BPBI S.%
\end{APACrefauthors}%
\unskip\
\newblock
\APACrefYearMonthDay{2023}{}{}.
\newblock
{\BBOQ}\APACrefatitle {Limitations of variational quantum algorithms: a quantum optimal transport approach} {Limitations of variational quantum algorithms: a quantum optimal transport approach}.{\BBCQ}
\newblock
\APACjournalVolNumPages{PRX Quantum}{4}{1}{010309}.
\PrintBackRefs{\CurrentBib}

\bibitem [\protect \citeauthoryear {%
De~Wolf%
}{%
De~Wolf%
}{%
{\protect \APACyear {2017}}%
}]{%
de2017potential}
\APACinsertmetastar {%
de2017potential}%
\begin{APACrefauthors}%
De~Wolf, R.%
\end{APACrefauthors}%
\unskip\
\newblock
\APACrefYearMonthDay{2017}{}{}.
\newblock
{\BBOQ}\APACrefatitle {The potential impact of quantum computers on society} {The potential impact of quantum computers on society}.{\BBCQ}
\newblock
\APACjournalVolNumPages{Ethics and Information Technology}{19}{}{271--276}.
\PrintBackRefs{\CurrentBib}

\bibitem [\protect \citeauthoryear {%
Fan%
\ \protect \BOthers {.}}{%
Fan%
\ \protect \BOthers {.}}{%
{\protect \APACyear {2025}}%
}]{%
fan2025calibrating}
\APACinsertmetastar {%
fan2025calibrating}%
\begin{APACrefauthors}%
Fan, D.%
, Liu, G.%
, Li, S.%
, Gong, M.%
, Wu, D.%
, Zhang, Y.%
\BDBL {}others%
\end{APACrefauthors}%
\unskip\
\newblock
\APACrefYearMonthDay{2025}{}{}.
\newblock
{\BBOQ}\APACrefatitle {Calibrating quantum gates up to 52 qubits in a superconducting processor} {Calibrating quantum gates up to 52 qubits in a superconducting processor}.{\BBCQ}
\newblock
\APACjournalVolNumPages{npj Quantum Information}{11}{1}{33}.
\PrintBackRefs{\CurrentBib}

\bibitem [\protect \citeauthoryear {%
Farhi%
, Goldstone%
\BCBL {}\ \BBA {} Gutmann%
}{%
Farhi%
\ \protect \BOthers {.}}{%
{\protect \APACyear {2014}}%
}]{%
farhi2014quantum}
\APACinsertmetastar {%
farhi2014quantum}%
\begin{APACrefauthors}%
Farhi, E.%
, Goldstone, J.%
\BCBL {}\ \BBA {} Gutmann, S.%
\end{APACrefauthors}%
\unskip\
\newblock
\APACrefYearMonthDay{2014}{}{}.
\newblock
{\BBOQ}\APACrefatitle {A quantum approximate optimization algorithm} {A quantum approximate optimization algorithm}.{\BBCQ}
\newblock
\APACjournalVolNumPages{arXiv preprint arXiv:1411.4028}{}{}{}.
\PrintBackRefs{\CurrentBib}

\bibitem [\protect \citeauthoryear {%
Javadi-Abhari%
\ \protect \BOthers {.}}{%
Javadi-Abhari%
\ \protect \BOthers {.}}{%
{\protect \APACyear {2024}}%
}]{%
javadi2024quantum}
\APACinsertmetastar {%
javadi2024quantum}%
\begin{APACrefauthors}%
Javadi-Abhari, A.%
, Treinish, M.%
, Krsulich, K.%
, Wood, C\BPBI J.%
, Lishman, J.%
, Gacon, J.%
\BDBL {}others%
\end{APACrefauthors}%
\unskip\
\newblock
\APACrefYearMonthDay{2024}{}{}.
\newblock
{\BBOQ}\APACrefatitle {Quantum computing with {Q}iskit} {Quantum computing with {Q}iskit}.{\BBCQ}
\newblock
\APACjournalVolNumPages{arXiv preprint arXiv:2405.08810}{}{}{}.
\PrintBackRefs{\CurrentBib}

\bibitem [\protect \citeauthoryear {%
Kim%
, Oh%
, Chong%
, Hwang%
\BCBL {}\ \BBA {} Park%
}{%
Kim%
\ \protect \BOthers {.}}{%
{\protect \APACyear {2022}}%
}]{%
kim2022quantum}
\APACinsertmetastar {%
kim2022quantum}%
\begin{APACrefauthors}%
Kim, J.%
, Oh, B.%
, Chong, Y.%
, Hwang, E.%
\BCBL {}\ \BBA {} Park, D\BPBI K.%
\end{APACrefauthors}%
\unskip\
\newblock
\APACrefYearMonthDay{2022}{}{}.
\newblock
{\BBOQ}\APACrefatitle {Quantum readout error mitigation via deep learning} {Quantum readout error mitigation via deep learning}.{\BBCQ}
\newblock
\APACjournalVolNumPages{New Journal of Physics}{24}{7}{073009}.
\PrintBackRefs{\CurrentBib}

\bibitem [\protect \citeauthoryear {%
Liao%
\ \protect \BOthers {.}}{%
Liao%
\ \protect \BOthers {.}}{%
{\protect \APACyear {2024}}%
}]{%
liao2024machine}
\APACinsertmetastar {%
liao2024machine}%
\begin{APACrefauthors}%
Liao, H.%
, Wang, D\BPBI S.%
, Sitdikov, I.%
, Salcedo, C.%
, Seif, A.%
\BCBL {}\ \BBA {} Minev, Z\BPBI K.%
\end{APACrefauthors}%
\unskip\
\newblock
\APACrefYearMonthDay{2024}{}{}.
\newblock
{\BBOQ}\APACrefatitle {Machine learning for practical quantum error mitigation} {Machine learning for practical quantum error mitigation}.{\BBCQ}
\newblock
\APACjournalVolNumPages{Nature Machine Intelligence}{}{}{1--9}.
\PrintBackRefs{\CurrentBib}

\bibitem [\protect \citeauthoryear {%
Lienhard%
\ \protect \BOthers {.}}{%
Lienhard%
\ \protect \BOthers {.}}{%
{\protect \APACyear {2022}}%
}]{%
lienhard2022deep}
\APACinsertmetastar {%
lienhard2022deep}%
\begin{APACrefauthors}%
Lienhard, B.%
, Veps{\"a}l{\"a}inen, A.%
, Govia, L\BPBI C.%
, Hoffer, C\BPBI R.%
, Qiu, J\BPBI Y.%
, Rist{\`e}, D.%
\BDBL {}others%
\end{APACrefauthors}%
\unskip\
\newblock
\APACrefYearMonthDay{2022}{}{}.
\newblock
{\BBOQ}\APACrefatitle {Deep-neural-network discrimination of multiplexed superconducting-qubit states} {Deep-neural-network discrimination of multiplexed superconducting-qubit states}.{\BBCQ}
\newblock
\APACjournalVolNumPages{Physical Review Applied}{17}{1}{014024}.
\PrintBackRefs{\CurrentBib}

\bibitem [\protect \citeauthoryear {%
Lubasch%
, Joo%
, Moinier%
, Kiffner%
\BCBL {}\ \BBA {} Jaksch%
}{%
Lubasch%
\ \protect \BOthers {.}}{%
{\protect \APACyear {2020}}%
}]{%
lubasch2020variational}
\APACinsertmetastar {%
lubasch2020variational}%
\begin{APACrefauthors}%
Lubasch, M.%
, Joo, J.%
, Moinier, P.%
, Kiffner, M.%
\BCBL {}\ \BBA {} Jaksch, D.%
\end{APACrefauthors}%
\unskip\
\newblock
\APACrefYearMonthDay{2020}{}{}.
\newblock
{\BBOQ}\APACrefatitle {Variational quantum algorithms for nonlinear problems} {Variational quantum algorithms for nonlinear problems}.{\BBCQ}
\newblock
\APACjournalVolNumPages{Physical Review A}{101}{1}{010301}.
\PrintBackRefs{\CurrentBib}

\bibitem [\protect \citeauthoryear {%
Maciejewski%
, Zimbor{\'a}s%
\BCBL {}\ \BBA {} Oszmaniec%
}{%
Maciejewski%
\ \protect \BOthers {.}}{%
{\protect \APACyear {2020}}%
}]{%
maciejewski2020mitigation}
\APACinsertmetastar {%
maciejewski2020mitigation}%
\begin{APACrefauthors}%
Maciejewski, F\BPBI B.%
, Zimbor{\'a}s, Z.%
\BCBL {}\ \BBA {} Oszmaniec, M.%
\end{APACrefauthors}%
\unskip\
\newblock
\APACrefYearMonthDay{2020}{}{}.
\newblock
{\BBOQ}\APACrefatitle {Mitigation of readout noise in near-term quantum devices by classical post-processing based on detector tomography} {Mitigation of readout noise in near-term quantum devices by classical post-processing based on detector tomography}.{\BBCQ}
\newblock
\APACjournalVolNumPages{Quantum}{4}{}{257}.
\PrintBackRefs{\CurrentBib}

\bibitem [\protect \citeauthoryear {%
Maksymov%
, Niroula%
\BCBL {}\ \BBA {} Nam%
}{%
Maksymov%
\ \protect \BOthers {.}}{%
{\protect \APACyear {2021}}%
}]{%
maksymov2021optimal}
\APACinsertmetastar {%
maksymov2021optimal}%
\begin{APACrefauthors}%
Maksymov, A.%
, Niroula, P.%
\BCBL {}\ \BBA {} Nam, Y.%
\end{APACrefauthors}%
\unskip\
\newblock
\APACrefYearMonthDay{2021}{}{}.
\newblock
{\BBOQ}\APACrefatitle {Optimal calibration of gates in trapped-ion quantum computers} {Optimal calibration of gates in trapped-ion quantum computers}.{\BBCQ}
\newblock
\APACjournalVolNumPages{Quantum Science and Technology}{6}{3}{034009}.
\PrintBackRefs{\CurrentBib}

\bibitem [\protect \citeauthoryear {%
Mavroeidis%
, Vishi%
, Zych%
\BCBL {}\ \BBA {} J{\o}sang%
}{%
Mavroeidis%
\ \protect \BOthers {.}}{%
{\protect \APACyear {2018}}%
}]{%
mavroeidis2018impact}
\APACinsertmetastar {%
mavroeidis2018impact}%
\begin{APACrefauthors}%
Mavroeidis, V.%
, Vishi, K.%
, Zych, M\BPBI D.%
\BCBL {}\ \BBA {} J{\o}sang, A.%
\end{APACrefauthors}%
\unskip\
\newblock
\APACrefYearMonthDay{2018}{}{}.
\newblock
{\BBOQ}\APACrefatitle {The impact of quantum computing on present cryptography} {The impact of quantum computing on present cryptography}.{\BBCQ}
\newblock
\APACjournalVolNumPages{arXiv preprint arXiv:1804.00200}{}{}{}.
\PrintBackRefs{\CurrentBib}

\bibitem [\protect \citeauthoryear {%
M{\"o}ller%
\ \BBA {} Vuik%
}{%
M{\"o}ller%
\ \BBA {} Vuik%
}{%
{\protect \APACyear {2017}}%
}]{%
moller2017impact}
\APACinsertmetastar {%
moller2017impact}%
\begin{APACrefauthors}%
M{\"o}ller, M.%
\BCBT {}\ \BBA {} Vuik, C.%
\end{APACrefauthors}%
\unskip\
\newblock
\APACrefYearMonthDay{2017}{}{}.
\newblock
{\BBOQ}\APACrefatitle {On the impact of quantum computing technology on future developments in high-performance scientific computing} {On the impact of quantum computing technology on future developments in high-performance scientific computing}.{\BBCQ}
\newblock
\APACjournalVolNumPages{Ethics and information technology}{19}{}{253--269}.
\PrintBackRefs{\CurrentBib}

\bibitem [\protect \citeauthoryear {%
Nachman%
, Urbanek%
, de Jong%
\BCBL {}\ \BBA {} Bauer%
}{%
Nachman%
\ \protect \BOthers {.}}{%
{\protect \APACyear {2020}}%
}]{%
nachman2020unfolding}
\APACinsertmetastar {%
nachman2020unfolding}%
\begin{APACrefauthors}%
Nachman, B.%
, Urbanek, M.%
, de Jong, W\BPBI A.%
\BCBL {}\ \BBA {} Bauer, C\BPBI W.%
\end{APACrefauthors}%
\unskip\
\newblock
\APACrefYearMonthDay{2020}{}{}.
\newblock
{\BBOQ}\APACrefatitle {Unfolding quantum computer readout noise} {Unfolding quantum computer readout noise}.{\BBCQ}
\newblock
\APACjournalVolNumPages{npj Quantum Information}{6}{1}{84}.
\PrintBackRefs{\CurrentBib}

\bibitem [\protect \citeauthoryear {%
Nation%
, Kang%
, Sundaresan%
\BCBL {}\ \BBA {} Gambetta%
}{%
Nation%
\ \protect \BOthers {.}}{%
{\protect \APACyear {2021}}%
}]{%
nation2021scalable}
\APACinsertmetastar {%
nation2021scalable}%
\begin{APACrefauthors}%
Nation, P\BPBI D.%
, Kang, H.%
, Sundaresan, N.%
\BCBL {}\ \BBA {} Gambetta, J\BPBI M.%
\end{APACrefauthors}%
\unskip\
\newblock
\APACrefYearMonthDay{2021}{}{}.
\newblock
{\BBOQ}\APACrefatitle {Scalable mitigation of measurement errors on quantum computers} {Scalable mitigation of measurement errors on quantum computers}.{\BBCQ}
\newblock
\APACjournalVolNumPages{PRX Quantum}{2}{4}{040326}.
\PrintBackRefs{\CurrentBib}

\bibitem [\protect \citeauthoryear {%
Pan%
, Chen%
\BCBL {}\ \BBA {} Zhang%
}{%
Pan%
\ \protect \BOthers {.}}{%
{\protect \APACyear {2022}}%
}]{%
pan2022solving}
\APACinsertmetastar {%
pan2022solving}%
\begin{APACrefauthors}%
Pan, F.%
, Chen, K.%
\BCBL {}\ \BBA {} Zhang, P.%
\end{APACrefauthors}%
\unskip\
\newblock
\APACrefYearMonthDay{2022}{}{}.
\newblock
{\BBOQ}\APACrefatitle {Solving the sampling problem of the sycamore quantum circuits} {Solving the sampling problem of the sycamore quantum circuits}.{\BBCQ}
\newblock
\APACjournalVolNumPages{Physical Review Letters}{129}{9}{090502}.
\PrintBackRefs{\CurrentBib}

\bibitem [\protect \citeauthoryear {%
Pokharel%
, Srinivasan%
, Quiroz%
\BCBL {}\ \BBA {} Boots%
}{%
Pokharel%
\ \protect \BOthers {.}}{%
{\protect \APACyear {2024}}%
}]{%
pokharel2024scalable}
\APACinsertmetastar {%
pokharel2024scalable}%
\begin{APACrefauthors}%
Pokharel, B.%
, Srinivasan, S.%
, Quiroz, G.%
\BCBL {}\ \BBA {} Boots, B.%
\end{APACrefauthors}%
\unskip\
\newblock
\APACrefYearMonthDay{2024}{}{}.
\newblock
{\BBOQ}\APACrefatitle {Scalable measurement error mitigation via iterative {B}ayesian unfolding} {Scalable measurement error mitigation via iterative {B}ayesian unfolding}.{\BBCQ}
\newblock
\APACjournalVolNumPages{Physical Review Research}{6}{1}{013187}.
\PrintBackRefs{\CurrentBib}

\bibitem [\protect \citeauthoryear {%
Preskill%
}{%
Preskill%
}{%
{\protect \APACyear {2019}}%
}]{%
preskill2019quantum}
\APACinsertmetastar {%
preskill2019quantum}%
\begin{APACrefauthors}%
Preskill, J.%
\end{APACrefauthors}%
\unskip\
\newblock
\APACrefYearMonthDay{2019}{}{}.
\newblock
{\BBOQ}\APACrefatitle {Quantum computing in the {NISQ} era and beyond} {Quantum computing in the {NISQ} era and beyond}.{\BBCQ}
\newblock
\APACjournalVolNumPages{Bulletin of the American Physical Society}{64}{}{9}.
\PrintBackRefs{\CurrentBib}

\bibitem [\protect \citeauthoryear {%
Rawat%
, Mehra%
, Bist%
, Yusup%
\BCBL {}\ \BBA {} Sanjaya%
}{%
Rawat%
\ \protect \BOthers {.}}{%
{\protect \APACyear {2022}}%
}]{%
rawat2022quantum}
\APACinsertmetastar {%
rawat2022quantum}%
\begin{APACrefauthors}%
Rawat, B.%
, Mehra, N.%
, Bist, A\BPBI S.%
, Yusup, M.%
\BCBL {}\ \BBA {} Sanjaya, Y\BPBI P\BPBI A.%
\end{APACrefauthors}%
\unskip\
\newblock
\APACrefYearMonthDay{2022}{}{}.
\newblock
{\BBOQ}\APACrefatitle {Quantum computing and {AI}: Impacts \& possibilities} {Quantum computing and {AI}: Impacts \& possibilities}.{\BBCQ}
\newblock
\APACjournalVolNumPages{ADI Journal on Recent Innovation}{3}{2}{202--207}.
\PrintBackRefs{\CurrentBib}

\bibitem [\protect \citeauthoryear {%
Resch%
\ \BBA {} Karpuzcu%
}{%
Resch%
\ \BBA {} Karpuzcu%
}{%
{\protect \APACyear {2021}}%
}]{%
resch2021benchmarking}
\APACinsertmetastar {%
resch2021benchmarking}%
\begin{APACrefauthors}%
Resch, S.%
\BCBT {}\ \BBA {} Karpuzcu, U\BPBI R.%
\end{APACrefauthors}%
\unskip\
\newblock
\APACrefYearMonthDay{2021}{}{}.
\newblock
{\BBOQ}\APACrefatitle {Benchmarking quantum computers and the impact of quantum noise} {Benchmarking quantum computers and the impact of quantum noise}.{\BBCQ}
\newblock
\APACjournalVolNumPages{ACM Computing Surveys (CSUR)}{54}{7}{1--35}.
\PrintBackRefs{\CurrentBib}

\bibitem [\protect \citeauthoryear {%
Saki%
, Alam%
\BCBL {}\ \BBA {} Ghosh%
}{%
Saki%
\ \protect \BOthers {.}}{%
{\protect \APACyear {2021}}%
}]{%
saki2021impact}
\APACinsertmetastar {%
saki2021impact}%
\begin{APACrefauthors}%
Saki, A\BPBI A.%
, Alam, M.%
\BCBL {}\ \BBA {} Ghosh, S.%
\end{APACrefauthors}%
\unskip\
\newblock
\APACrefYearMonthDay{2021}{}{}.
\newblock
{\BBOQ}\APACrefatitle {Impact of noise on the resilience and the security of quantum computing} {Impact of noise on the resilience and the security of quantum computing}.{\BBCQ}
\newblock
\BIn{} \APACrefbtitle {2021 22nd International Symposium on Quality Electronic Design (ISQED)} {2021 22nd international symposium on quality electronic design (isqed)}\ (\BPGS\ 186--191).
\PrintBackRefs{\CurrentBib}

\bibitem [\protect \citeauthoryear {%
Stilck~Fran{\c{c}}a%
\ \BBA {} Garcia-Patron%
}{%
Stilck~Fran{\c{c}}a%
\ \BBA {} Garcia-Patron%
}{%
{\protect \APACyear {2021}}%
}]{%
stilck2021limitations}
\APACinsertmetastar {%
stilck2021limitations}%
\begin{APACrefauthors}%
Stilck~Fran{\c{c}}a, D.%
\BCBT {}\ \BBA {} Garcia-Patron, R.%
\end{APACrefauthors}%
\unskip\
\newblock
\APACrefYearMonthDay{2021}{}{}.
\newblock
{\BBOQ}\APACrefatitle {Limitations of optimization algorithms on noisy quantum devices} {Limitations of optimization algorithms on noisy quantum devices}.{\BBCQ}
\newblock
\APACjournalVolNumPages{Nature Physics}{17}{11}{1221--1227}.
\PrintBackRefs{\CurrentBib}

\bibitem [\protect \citeauthoryear {%
Sun%
\ \BBA {} Zhang%
}{%
Sun%
\ \BBA {} Zhang%
}{%
{\protect \APACyear {2013}}%
}]{%
sun2013sparse}
\APACinsertmetastar {%
sun2013sparse}%
\begin{APACrefauthors}%
Sun, T.%
\BCBT {}\ \BBA {} Zhang, C\BHBI H.%
\end{APACrefauthors}%
\unskip\
\newblock
\APACrefYearMonthDay{2013}{}{}.
\newblock
{\BBOQ}\APACrefatitle {Sparse matrix inversion with scaled {L}asso} {Sparse matrix inversion with scaled {L}asso}.{\BBCQ}
\newblock
\APACjournalVolNumPages{The Journal of Machine Learning Research}{14}{1}{3385--3418}.
\PrintBackRefs{\CurrentBib}

\bibitem [\protect \citeauthoryear {%
Svastits%
\ \protect \BOthers {.}}{%
Svastits%
\ \protect \BOthers {.}}{%
{\protect \APACyear {2025}}%
}]{%
svastits2025readout}
\APACinsertmetastar {%
svastits2025readout}%
\begin{APACrefauthors}%
Svastits, D.%
, Het{\'e}nyi, B.%
, Sz{\'e}chenyi, G.%
, Wootton, J.%
, Loss, D.%
, Bosco, S.%
\BCBL {}\ \BBA {} P{\'a}lyi, A.%
\end{APACrefauthors}%
\unskip\
\newblock
\APACrefYearMonthDay{2025}{}{}.
\newblock
{\BBOQ}\APACrefatitle {Readout sweet spots for spin qubits with strong spin-orbit interaction} {Readout sweet spots for spin qubits with strong spin-orbit interaction}.{\BBCQ}
\newblock
\APACjournalVolNumPages{arXiv preprint arXiv:2505.15878}{}{}{}.
\PrintBackRefs{\CurrentBib}

\bibitem [\protect \citeauthoryear {%
Yang%
, Raymond%
\BCBL {}\ \BBA {} Uno%
}{%
Yang%
\ \protect \BOthers {.}}{%
{\protect \APACyear {2022}}%
}]{%
yang2022efficient}
\APACinsertmetastar {%
yang2022efficient}%
\begin{APACrefauthors}%
Yang, B.%
, Raymond, R.%
\BCBL {}\ \BBA {} Uno, S.%
\end{APACrefauthors}%
\unskip\
\newblock
\APACrefYearMonthDay{2022}{}{}.
\newblock
{\BBOQ}\APACrefatitle {Efficient quantum readout-error mitigation for sparse measurement outcomes of near-term quantum devices} {Efficient quantum readout-error mitigation for sparse measurement outcomes of near-term quantum devices}.{\BBCQ}
\newblock
\APACjournalVolNumPages{Physical Review A}{106}{1}{012423}.
\PrintBackRefs{\CurrentBib}

\end{thebibliography}

\end{document}